\documentclass[aps,pre,twocolumn,superscriptaddress,showpacs,showkeys ]{revtex4-2}

\DeclareUnicodeCharacter{2061}{}
\usepackage{graphicx}                   
\usepackage{hyperref}                   
\usepackage{amsmath,amssymb}            
\usepackage{epstopdf}
\usepackage{subcaption}	      			

\usepackage{color}
\definecolor{orange}{rgb}{1,0.5,0}
\definecolor{brown}{rgb}{0.65, 0.16, 0.16}
\definecolor{phlox}{rgb}{0.87, 0.0, 1.0}

\graphicspath{{figs/}}					

\begin{document}
   \title{Mapping cumulus clouds to scale invariant rough surfaces}
	\author{J. Cheraghalizadeh}
	\affiliation{Department of Physics, University of Mohaghegh Ardabili, P.O. Box 179, Ardabil, Iran}
     \author{S. Tizdast}
     \affiliation{Department of Physics, University of Mohaghegh Ardabili, P.O. Box 179, Ardabil, Iran}
    \author{H. Mohammadzade}
   \affiliation{Department of Physics, University of Mohaghegh Ardabili, P.O. Box 179, Ardabil, Iran}
    \author{M. N. Najafi}
    \affiliation{Department of Physics, University of Mohaghegh Ardabili, P.O. Box 179, Ardabil, Iran}
    \email{morteza.nattagh@gmail.com}
\begin{abstract}	
Motivated by a recent observation on the self-organized criticality of cumulus clouds (Phys. Rev E 103, 052106, 2021) we study their connection to self-similar rough surfaces, in which $f\equiv \log I$ plays the role of the main field, where $I$ is the intensity of the received visible light. By simulating the light scattering based on a coarse-grained phenomenological model in a two-dimensional cloud, we argue the possible connection of $I$ to the actual cloud thickness. Although in the vertical incident light $f$ is proportional to the cloud thickness, in the general case it is complected. We study the statistical properties of observational data for $f$ with a focus on the conventional exponents of this scale-invariant rough surface. By calculating the roughness exponents, and comparing them with other exponents like the fractal dimension of loops, the distribution function of the radius of gyration and loop lengths, and the exponent of the green function, we prove that this surface is unconventional in the sense that it is the non-Gaussian self-affine random surface which violates the Kondev hyper-scaling relations. 
\end{abstract}
\maketitle
	
\section{Introduction}
Due to their influence in meteorology, climate and water resources, clouds have been the source of much studies. Multi-fractality of clouds was proven directly~\cite{najafi2021self,lovejoy1982area,austin1985small,malinowski1993surface,lovejoy1991multifractal,lovejoy1987functional,cahalan1989fractal,gabriel1988multifractal,hentschel1984relative,sanchez2005fractal} and indirectly (e.g. scaling relation between rainfall environment and cloud area)~\cite{lovejoy1982area,pelletier1997kardar,rys1986fractal}, based on which attempts have been made to classify them in terms of their statistical properties~\cite{lovejoy1990multifractals,tessier1993universal,pelletier1997kardar}, or their morphology\cite{sengupta1990cumulus}, and also a few models appeared to describe their properties.
These models are divided into three categories, which are: turbulence-based models, cellular automaton models, and phenomenological models. For a good review see~\cite{najafi2021self}, which claims that they show self-organized criticality, in much accordance with the sandpile models. This property has also been observed for precipitation time series~\cite{peters2006critical} and has been obtained by analyzing ground-to-sky images~\cite{najafi2021self}. Fractality and criticality of cumulus clouds are two important ingredients, the exact mechanism behind of which has been a mystery, the understanding of which needs a detailed analysis of their observational. These two ingredients motivate one to map them to rough random surfaces, where the (logarithm of the) received intensity plays the role of the fluctuating random field, the curiosity that has already been addressed using the KPZ universality class~\cite{pelletier1997kardar}. In this point of view, the roughness (as an important factor in the reflection and absorption of solar energy from the clouds~\cite{thekkekara2017bioinspired,twomey1967light}) and also the distribution of the logarithm of the received light intensity are the dominant leading parameters to be analyzed in the present paper. Identifying the universality class of such a complex system helps much to realize the properties of the system in terms of simpler systems, see for example its relation to the sandpile models~\cite{najafi2021self,najafi2020geometry}. We present a systematic study on this system using the concepts in scale-invariant rough surfaces. \\

Random surfaces are widely used in physics to model phenomena at various scales, from the growth of rough surfaces at the nano- and micro-scales to very large systems at the cosmic scale. They also describe the fracture in materials science \cite{bouchaud1990fractal,bouchaud1993statistics}, Ripple-Wave Turbulence \cite{wright1997imaging} and passive detectors in the flow of two-dimensional fluids \cite{ramshankar1991transport,cardoso1996dispersion}. There are many universality classes for two-dimensional scale-invariant rough surfaces, like the Edwards-Wilkinson~\cite{hosseinabadi2013universality}, wolf-villain~\cite{wolf1990growth}, Kardar-Parzi-Zhang (KPZ)~\cite{pelletier1997kardar}. Many other phenomena are related to rough surfaces, like the depinning transition, where exactly the same concepts are used~\cite{valizadeh2021edwards}. A large class of rough surfaces is called Gaussian surfaces, for which the distribution function of the fluctuating field is Gaussian, and the exponent of the loop Green function is conjectured to be $x_l=\frac{1}{2}$~\cite{kondev2000nonlinear}. Other electronic system examples are the electron-hole puddles in Graphene~\cite{najafi2016conformal,najafi2018scaling} superconductivity in YBCO compounds~\cite{najafi2016universality} and Gaussian free fields~\cite{cheraghalizadeh2018gaussian1,cheraghalizadeh2018gaussian2} are known to be non-Gaussian. For a good review see~\cite{wu1988non}. The relation of the cumulus clouds to the rough surfaces was partially considered in~\cite{Cheraghalizade2020cloud}. The present article is devoted to a comprehensive study of the relationship between the 2D intensity field of cumulus clouds and rough surfaces and is divided into two parts. In the first part, we present a coarse-grained phenomenological model based on scattering, using which we show that the cloud thickness is proportional to the logarithm of the received light intensity. In the second part, we examine the statistical properties of the scattered light intensity and analyze various exponents, like the roughness exponent and the distribution function of the intensity filed. Comparing the results with the previous exponent, we show that the Kondev hyperscaling relations do no hold.\\

The paper is organized as follows: in the next section we introduce the main problem and describe briefly the standard theory of scale-invariant rough surfaces. Section~\ref{SEC:observation} is devoted to a description of the observational data and the results, with a focus on obtaining the exponents. We describe the simulation of light scattering of cumulus clouds in SEC.~\ref{SEC:simulations}. We finally describe the Kondev hyper-scaling relations in SEC.~\ref{SEC:Kondev}, and close the paper by a conclusion.
	
\section{Intensity field as a standard scale-invariant Rough surface}
In this paper, we aim to analyze two-dimensional maps of the intensities recorded from the cumulus clouds. For a given received intensity $I$, we consider
\begin{equation}
f(\textbf{r}) = \log I(\textbf{r})
\label{eq:intensity}
\end{equation}
as the fluctuating field, defining a rough surface, where $I(\textbf{r})$ shows the intensity recorded at the position $\textbf{r}$ in a 2D map, i.e. a particular pixel in the taken photo. This intensity is scaled automatically by the camera to a number between zero and $256$. From this definition, one observes that under the transformation $I\rightarrow \lambda I$ ($\lambda$ being a positive scaling factor), $f\rightarrow f + \log \lambda$, from which we see that an arbitrary rescaling can affect the properties of the system.  For the future convenience we fix the scale $\lambda$ by $f \to f-f_0$, where ${f_0} = \log I_0$ and $I_0$ are the average quantities over each sample.\\

In the standard theory of self-similar (more precisely self-affine) rough surfaces, the patterns repeat in all scales, meaning that their behaviors under the spatial re-scaling do not change. The the self affinity of the fluctuating field (in a spatial point $\textbf{r}$) $f(\textbf{r})$ is expressed by means of the relation
\begin{equation}
f(\textbf{r}) \overset{d}{=} \lambda^{ - \alpha }f(\lambda_r\textbf{r}) 
\label{Eq:scaleFree}
\end{equation}
where $\lambda_r$ is a positive re-scaling number. In this relation $\overset{d}{=}$ means the equality of the distribution functions, and $\alpha$ is called the \textit{roughness} exponent. A wide class of rough surfaces called Gaussian surfaces are expressed by the following distribution function\cite{kondev2000nonlinear}:
\begin{equation}
P\{ f\}  \sim \exp \left[ {\frac{{ - \kappa }}{2}\int_0^{\Lambda}  {{q^{2(1 + \alpha )}}\tilde{f}(q)\tilde{f}( - q){\text{d}^2}q} } \right],
\label{Eq:pattern}
\end{equation}
where $\tilde{f}(q)$ is the Fourier transform of $f (\textbf{r})$, $\Lambda$ is a cutoff momentum and $\kappa$ is the stiffness parameter. This serves as a formal definition, while for a more practical investigation one needs some other statistical observables. Arguably the roughness is the most important quantity in characterizing rough surfaces, defined as
\begin{equation}
W(L) \equiv \overline{\left[ {f(\textbf{r}) - \overline{f}} \right]^2}
\label{Eq:roughness}
\end{equation}
where the overline (like $ \bar f$) is the spatial average of the fluctuating field over the sample (inside a box of linear length $L$). To avoid noise, one has to take an ensemble average for Eq.~\ref{Eq:roughness} (from now on we show ensemble averaging by the symbol $\left\langle \right\rangle $). For scale-invariant rough surfaces, this function behaves like $ {L^{ - 2{\alpha_g}}}$, where the parameter $\alpha_{g}$ is called the \textit{global roughness exponent}, which should be less than one~\cite{barabasi1995fractal}. Another roughness exponent comes from the height-height correlation function ($\textbf{r}_0$ is a reference point)
\begin{equation}
C(\textbf{r}) \equiv \left\langle \left[f(\textbf{r}+\textbf{r}_0)-f(\textbf{r}_0)\right]^2 \right\rangle
\label{Eq:correlation}
\end{equation}
which behaves like $r^{-2\alpha_l}$, where $\alpha_{l}$ is called the \textit{local roughness exponent} \cite{barabasi1995fractal}. For mono-fractal scale-free systems these two exponents are the same, i.e. $\alpha_{l} = \alpha_{g} = \alpha$~\cite{kondev2000nonlinear}. \\

Valuable information about the rough surfaces can be obtained by Gaussianity tests showing whether or not a surface is Gaussian, including the distribution function of the fluctuating field $P(f)$, and the local curvature. In fact, a wide class of rough surfaces are Gaussian random surfaces in which $P(f)$ is Gaussian. The local curvature is defined in position $\textbf{r}$ and scale $b$ as follows \cite{kondev2000nonlinear}
\begin{equation}
C_b(\textbf{r}) = \sum_{m = 1}^M \left( f(\textbf{r} + b\textbf{e}_m) - f(\textbf{r})\right).
\label{6}
\end{equation}
In this relation, the set of directions $ \left\{ \textbf{e}_1,...,\textbf{e}_M \right\}$ are fixed single vectors whose sum is zero. If the surface is Gaussian roughness, the distribution function of $C_b$ would be Gaussian. As a test for Gaussianity of a distribution, the kurtosis should be examined
\begin{equation}
F_b = \frac{{\left\langle {C_b^4} \right\rangle }}{{{{\left\langle {C_b^2} \right\rangle }^2}}}
\label{Eq:Biner} 
\end{equation}
which is $3$ for a Gaussian distribution. The symmetries of a random surfaces also give us valuable information about the universality class of the system, one of which is reflection symmetry $(f(\textbf{r}) \leftrightarrow  - f(\textbf{r})) $. If a system is invariant under such a transformation, then it is expected that all of the odd moments will be zero ($\left\langle {C_b^q} \right\rangle  =0$ for odd $q$ values). \\

One of the most important theoretical studies of iso-lines (contours) of two-dimensional scale-invariant rough surfaces was coined by Kondev~\cite{kondev2000nonlinear}, who conjectured some relations between statistical properties of level lines of a rough surface and the exponents explored above. The level lines are defined as the locus of points satisfying $f(r)=f_0$. Then one considers different levels between the maximum and minimum of $f$ with equal spacing ($10$ in the present paper), the total set of lines forms the contour loop ensemble (CLE), which is statistically investigated. If we show the distribution functions by $P(x)$, where $x$ is the loop length $l$, and the loop gyration radius $r_l$, then it has already been shown that 
\begin{equation}
    P(x)\sim x^{-\tau_x}
\end{equation}
where $\tau_x$ are the corresponding exponents. Also, the contour loop correlation function, which represents the probability that two points being separated by the distance $r$ are in the same in a contour line (loop), is shown to behave like 
\begin{equation}
G(r) \sim \frac{1}{{{r^{2{x_l}}}}} 
\end{equation}
for large $r$s, where $x_l$ is the loop correlation exponent, being a super-universal $1/2$ for Gaussian rough surfaces as was conjectured by Kondev~\cite{kondev2000nonlinear}. Additionally, $l$ and $r$ show scaling relation as follows 
\begin{equation}
\left\langle l \right\rangle  \sim {r^{{D_f}}} 
\end{equation}
where $D_f$ is the loop fractal dimension. It was conjectured by Kondev that there are some hype-scaling relations between the exponents explored above. Two important hyper-scaling relations are between $\tau_r$, $\tau_l$, $D_f$ and $\alpha$ as follows~\cite{kondev2000nonlinear}
\begin{equation}
\tau_r=1+D_f(\tau_l-1)=3-\alpha.
\label{Eq:KondevTau_r}
\end{equation}
Combining these relations, one finds that
\begin{equation}
D_f({\tau_l} - 1)  = 2 - \alpha.
\label{Eq:KondevTau_l}
\end{equation}
There is also another hyper-scaling relation that connects the exponents to $x_l$ as follows 
\begin{equation}
D_f({\tau _l} - 3)  = 2{x_l} - 2.
\label{Eq:KondevX_l}
\end{equation}
For the cases where the conditional probability $p(l|r)$ is small, one has $p(l)\text{d}l=p(r)\text{d}r$, which when is combined with $l\propto r^{D_f}$, results to the hyper-scaling relation
\begin{equation}
D_f  = \frac{{{\tau _l} - 1}}{{{\tau _r} - 1}}.
\label{Eq:KondevHyper}
\end{equation}
Also, the cumulative distribution of the number of contours with an area greater than A is another interesting quantity with the scaling property. The cumulative distribution of area $N_{>}(A)$ has the scaling form
\begin{equation}
N_{>}(A) \sim N^{-\xi/2}
\end{equation}
where $\xi$ is a new exponent which is related to the roughness exponent by 
\begin{equation}
    \xi=2-\alpha.
\label{Eq:Hyper2}
\end{equation} 
These hype-scaling relations are investigated in this paper for the rough intensity field of the clouds.

\section{Light intensity observations}\label{SEC:observation}
\begin{figure*}
	\begin{subfigure}{0.47\textwidth}\includegraphics[width=\textwidth]{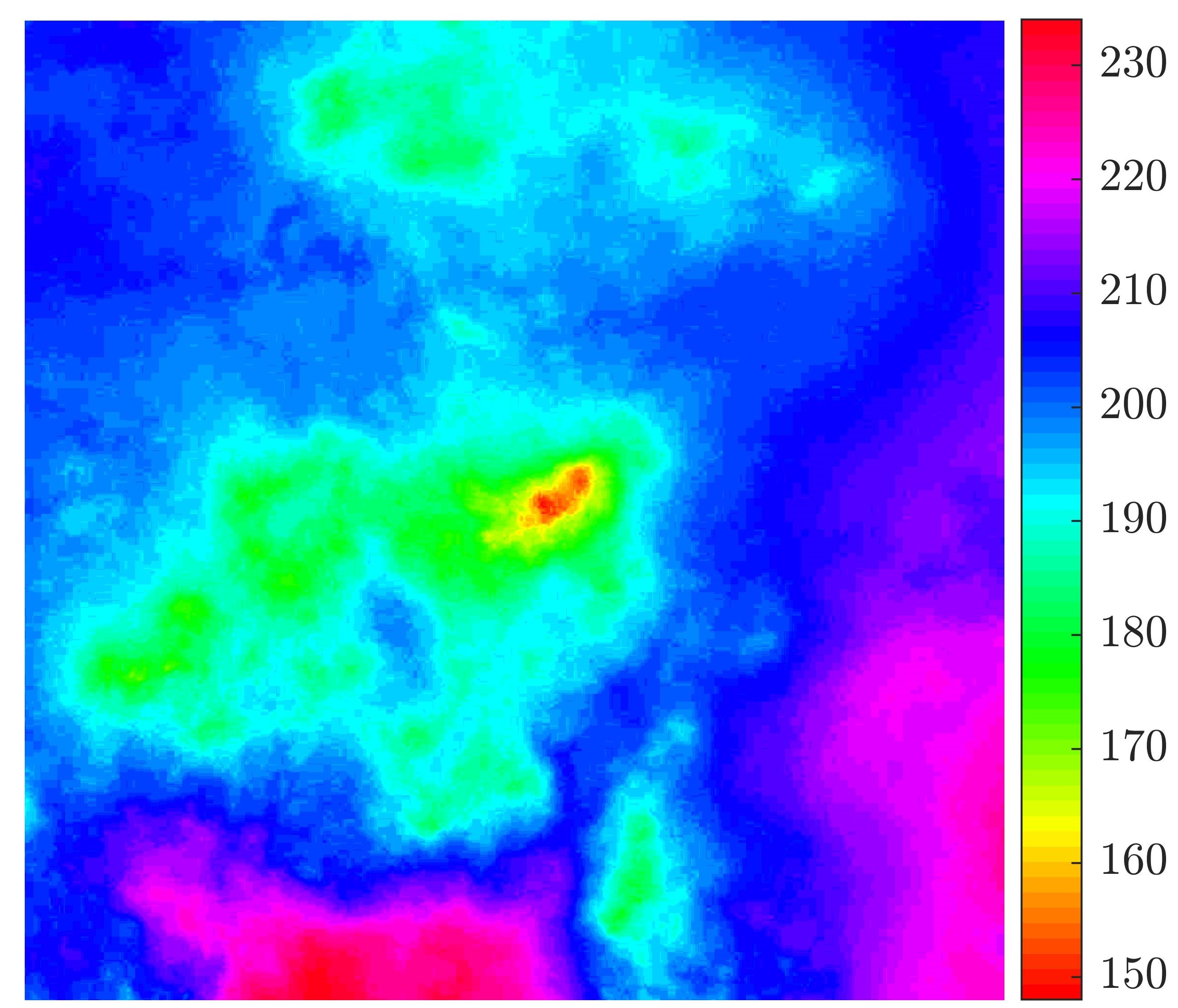}
		\caption{}
		\label{fig:fig2a}
	\end{subfigure}
	\begin{subfigure}{0.46\textwidth}\includegraphics[width=\textwidth]{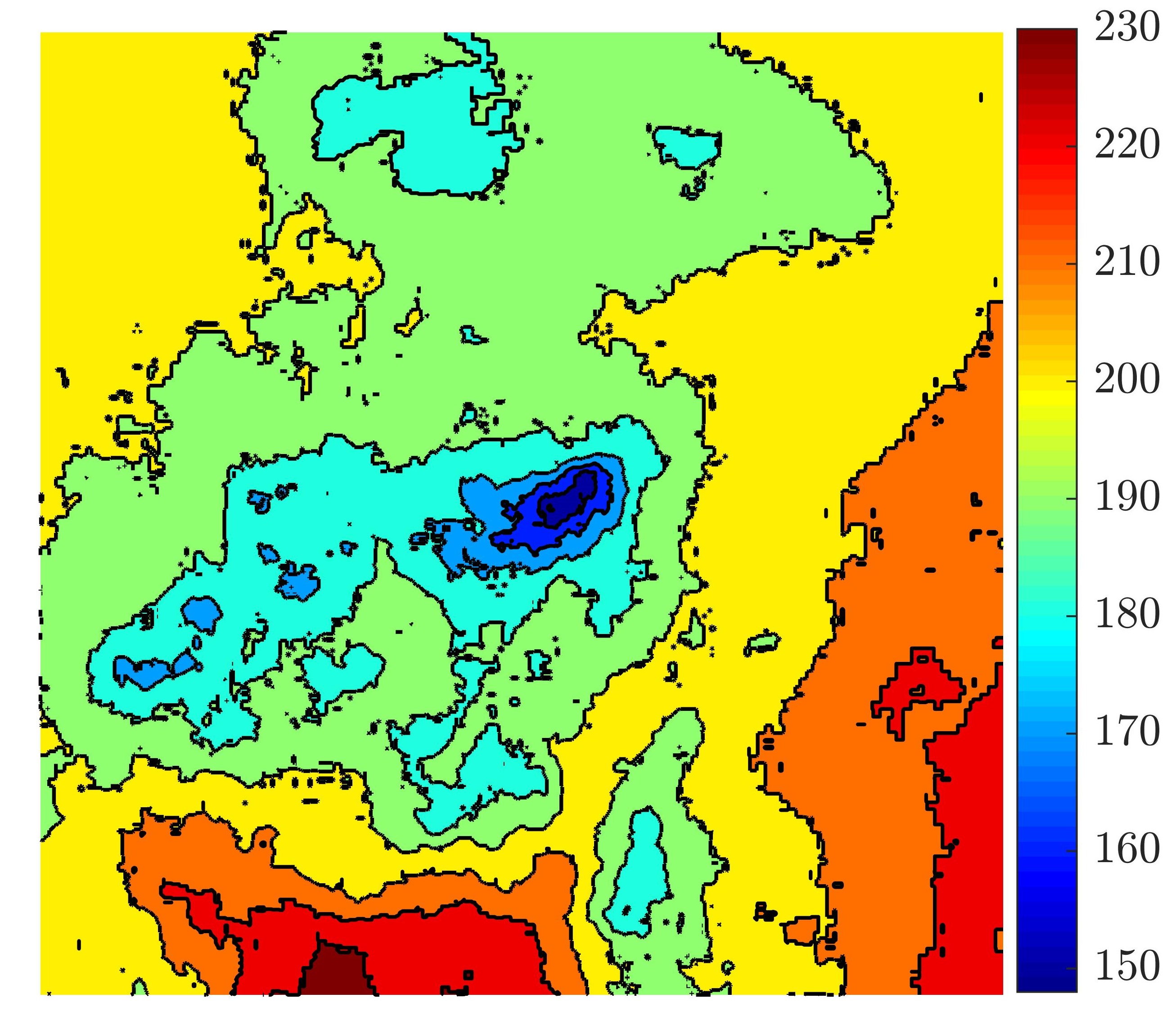}
		\caption{}
		\label{fig:fig2b}
	\end{subfigure}
	\caption{(a) Color map of the intensity recived from on sample of cumulus clouds, and (b) the corresponding contour plot. }
\end{figure*}

There are two main sources of light coming from the clouds: the sky background blue light (reaching the cloud from all directions), and the direct light from the sun. The light is scattered inside the clouds by the water droplets or the ice grains and also the aerosol/dust particles~\cite{plass1971radiative}. The cross-section of the visible light scattering depends on the radius distribution of these particles, which is mainly considered to be a gamma distribution~\cite{levin1958functions}. Generally, if the size of scatterers is much smaller than the incident beam wavelength, the Rayleigh scattering theory is used, while in the opposite case the geometrical optics is applicable. For the case where the particle size is comparable with the wavelength, the Mie scattering theory governs, in which the scattering cross-section is proportional to $\lambda ^ {-2}$. This latter is the case for light scattering from cumulus clouds~\cite{plass1968monte}. \\
In this section, we describe the statistics of the received light from the cumulus clouds. To this end, we analyzed over $100$ photographs with $"1000 \times 1000"$ pixels taken from cumulus clouds using a Nikon d7200 camera. The photos were taken more or less in the same conditions, i.e. at the same hours of the days with similar weather conditions and small wind strengths, in May and June of 2019 in Ardabil, Iran. Temporal and spatial details of the photographs, the angle of the sun with the ground at the time of photography, and the weather conditions are shown in Table 1 of ref.~\cite{Cheraghalizade2020cloud}. The received intensities were converted to (RGB) color maps, and then converted to gray-scale maps. Assuming that the distance between the cumulus clouds and the camera is about $2000$ meters, the size of each cloud piece that is recorded in a camera pixel $(3.9 \times 3.9\mu {m^2})$ is about $16 \times 16cm ^ 2$. Fig.~\ref{fig:fig2a} shows an example of the received intensity, the contour plots of which are shown in Fig.~\ref{fig:fig2b}. Before presenting the statistical properties of the received intensities, it is tempting to simulate the light in the clouds and find a relation between the received light and the thickness of the cloud, which the subjective of the next section.

\section{Simulation of light scattering from the cloud}\label{SEC:simulations}

	\begin{figure*}
	\begin{subfigure}{0.30\textwidth}\includegraphics[width=\textwidth]{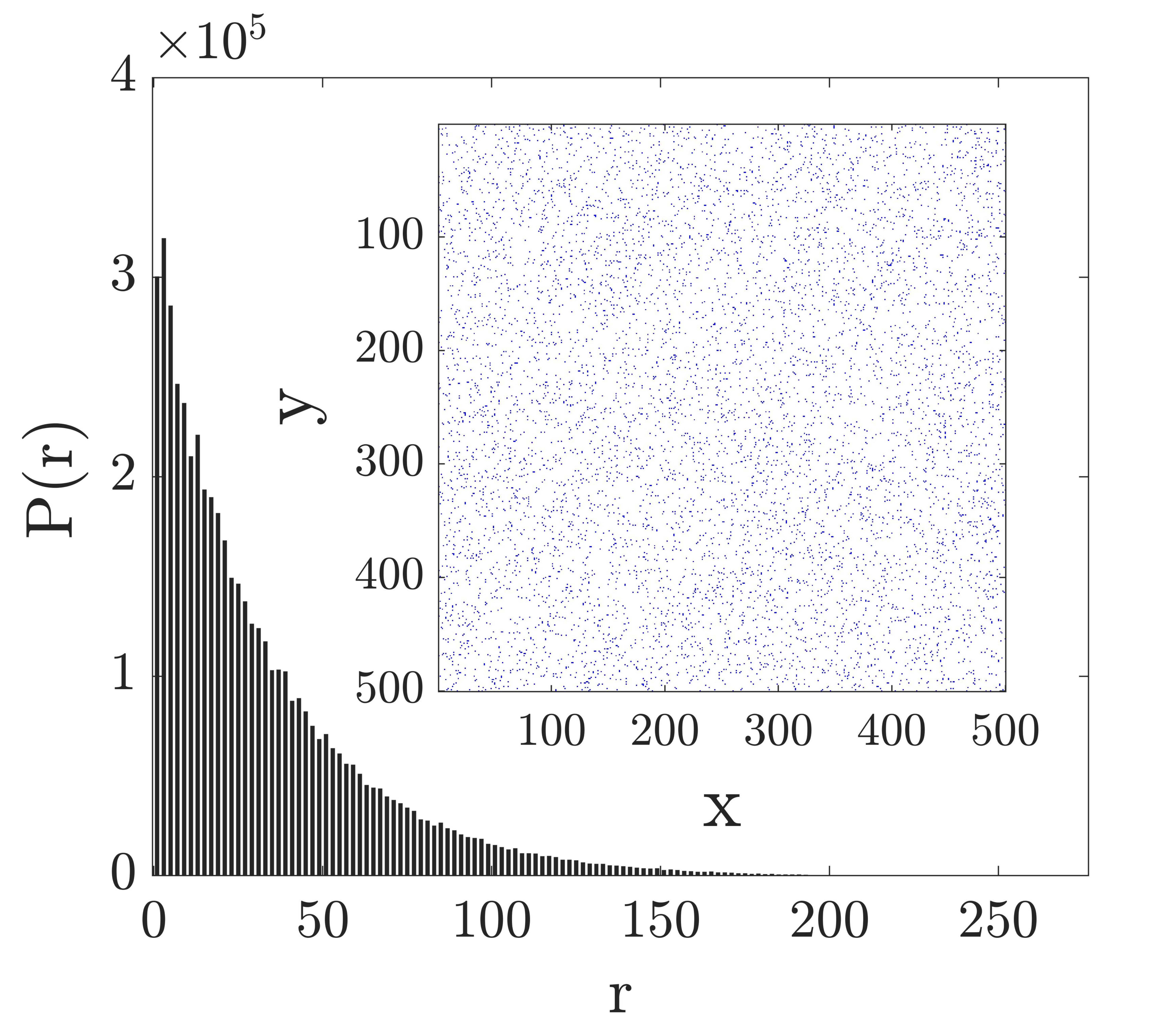}
		\caption{}
		\label{fig:lightscatteringRandom}
	\end{subfigure}
	\begin{subfigure}{0.33\textwidth}\includegraphics[width=\textwidth]{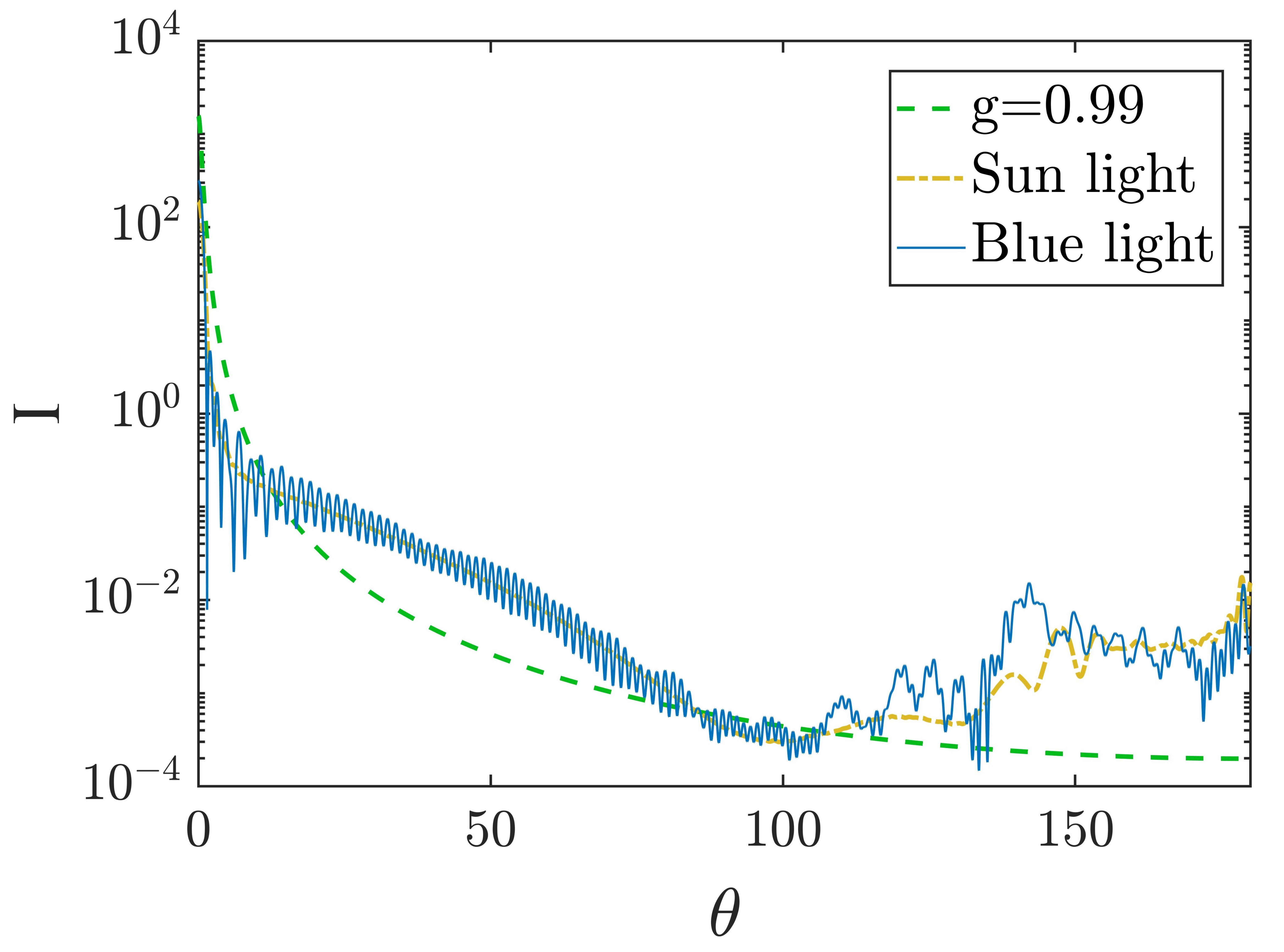}
		\caption{}
		\label{fig:lightscatteringMie}
	\end{subfigure}
	\begin{subfigure}{0.33\textwidth}\includegraphics[width=\textwidth]{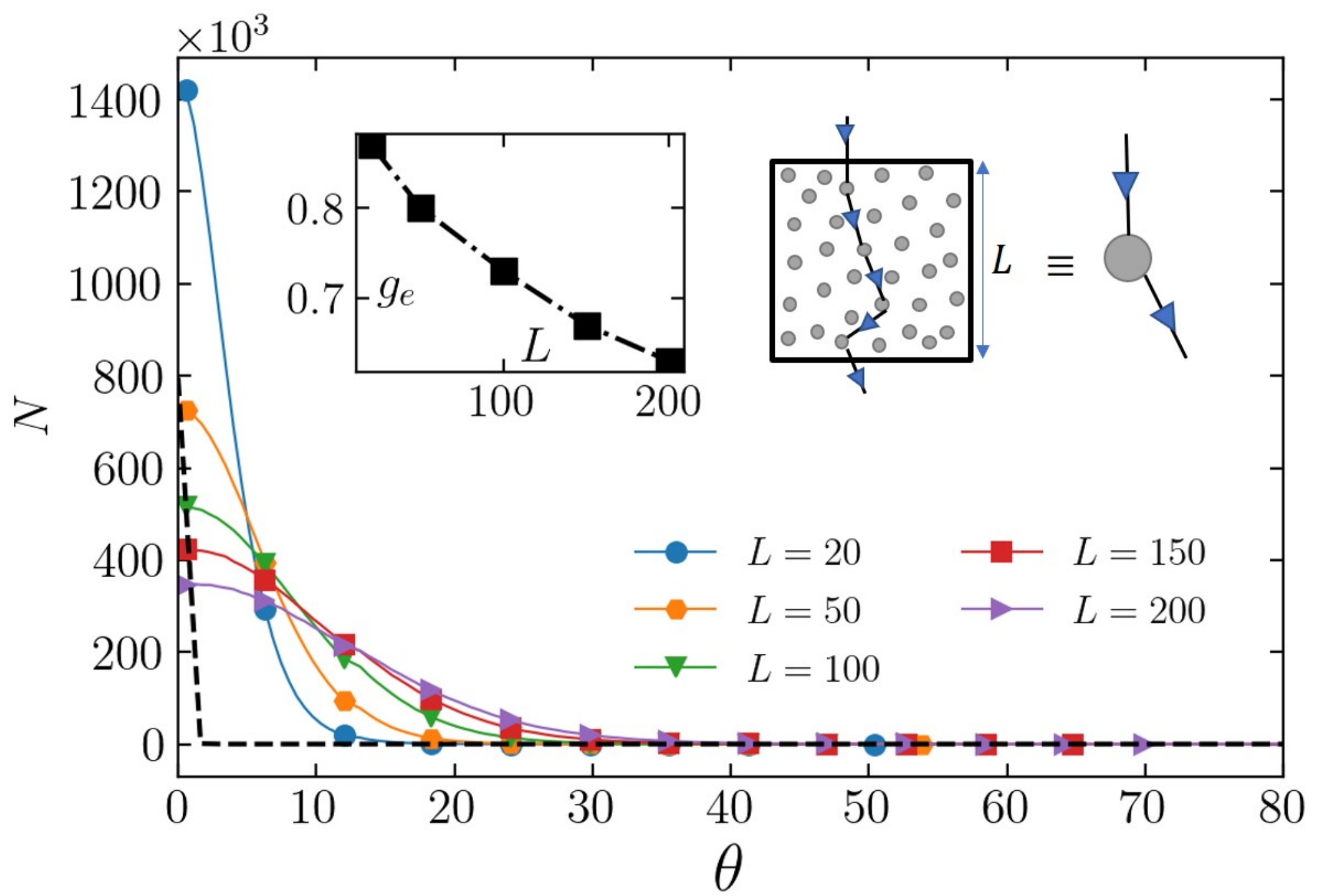}
		\caption{}
		\label{fig:lightscatteringEffective}
	\end{subfigure}
	\caption{The distribution of free path of light beams inside a cloud. Inset: the corresponding sample with random-positioned water droplets. (b) Mie scattering intensity in terms of scattering angle for sunlight and blue light. The Henny-Greenstein theory (Eq.~\ref{Eq:P}) with $g = 0.99$ is also shown. (c) Angle distribution of light scattered from cumulus clouds, and the fitting with the Henny-Greenstein theory. The effective $g_e$ is shown in the left inset, while in the right inset the setup has been shown.}
	\label{fig:lightscattering}
\end{figure*}
This section is devoted to exploring the coarse-grained model developed in{\color{red}~\cite{Cheraghalizade2020cloud}} for the light scattering in cumulus clouds, in order to understand the possible connection between the received light and the cloud thickness. Given that the mentioned paper was written in Farsi, we explain the model in detail here to be convenient for general audiences. This section can be skipped without harming the flow of the paper.\\

As explained in the previous section, the scattering of light inside the cumulus clouds follows the Mie scattering theory. Aside from the Mie scattering of visible light, the complexity of the problem arises from the multiple scattering due to the huge number of scatterers (water droplets), which is like a directional correlated random walk with stochastic step sizes~\cite{Cheraghalizade2020cloud}. In such a situation, one has to down-fold the system, to a system with a lower number of degrees of freedom by coarse-graining. In this approach, the system is supposed to be composed of many air parcels each of which has a huge number of water droplets and at the same time small enough in order to be treated as an infinitesimal in a large scale system. In Fig.~\ref{fig:lightscattering} we show the main ingredients of this model. Consider a two-dimensional system with a given density, inside which the water droplets are positioned uniformly at random according to Fig.~\ref{fig:lightscatteringRandom} (with the density $0.25 \ \text{drops} / mm ^ 2$, note that the average distance between particles in cumulus clouds is two millimeters \cite{durbin1959droplet}), where the distribution function of the distance between droplets is shown in the main part. The Mie scattering of a single droplet (with the radius $r=10 mm$) is shown in Fig.~\ref{fig:lightscatteringMie} for the sunlight and the sky blue light, where the dashed line shows the fitting with the following Henny-Greenstein function~\cite{max1995efficient} 
\begin{equation}
	P(\theta ) \propto \frac{{1 - {g^2}}}{{{{(1 + {g^2} - 2g\cos (\theta ))}^{\frac{3}{2}}}}},
	\label{Eq:P}
\end{equation}
where $\theta$ is the scattering angle. This function is generally used as an approximation of light scattering from the cumulus clouds~\cite{max1995efficient}. Sometimes other functions such as the Gaussian function~\cite{premovze2004practical} are also used, which is not as effective as the Henny-Greenstein function. Note that when $g=0$ we have random scattering (the scattering probability to all $\theta$ directions is the same), and when $g=1$ we only have forward scattering ($P$ is non-zero only for $\theta=0$, i.e. the light passes the scatterer without changing the direction). From Fig.~\ref{fig:lightscatteringMie} we see a good agreement with $g=0.99$, but we should have in mind that it is for single scattering, while light entering a cumulus cloud faces a lot of scattering centers, making $g$ effectively smaller. To show this we calculate the effective angle distribution function for the system with linear size $L$ composed of a large number of scatters, and calculate the effective $g{e}$ by fitting with Eq.~\ref{Eq:P}. The result is shown in Fig.~\ref{fig:lightscatteringEffective}, from which we observe that as $L$ increases, the scattering distribution becomes more uniform and the probability of scattering to larger angles increases, and eventually $g_e$ goes to zero as $L\rightarrow\infty$ so that a cumulus cloud is white as seen from earth~\cite{bouthors2008interactive}. In the latter limit, the light path inside the cumulus clouds becomes much like a non-directional random walk with random step sizes. For simulations, however, one needs air parcels with finite sizes for which $g_e$ is a non-zero quantity. In particular, in Ref.~\cite{Cheraghalizade2020cloud} the minimal box size $L_{\text{min}} = 20mm$ was used for which $g_e\approx0.85$.
\begin{figure}
	\centerline{\includegraphics[scale=.35]{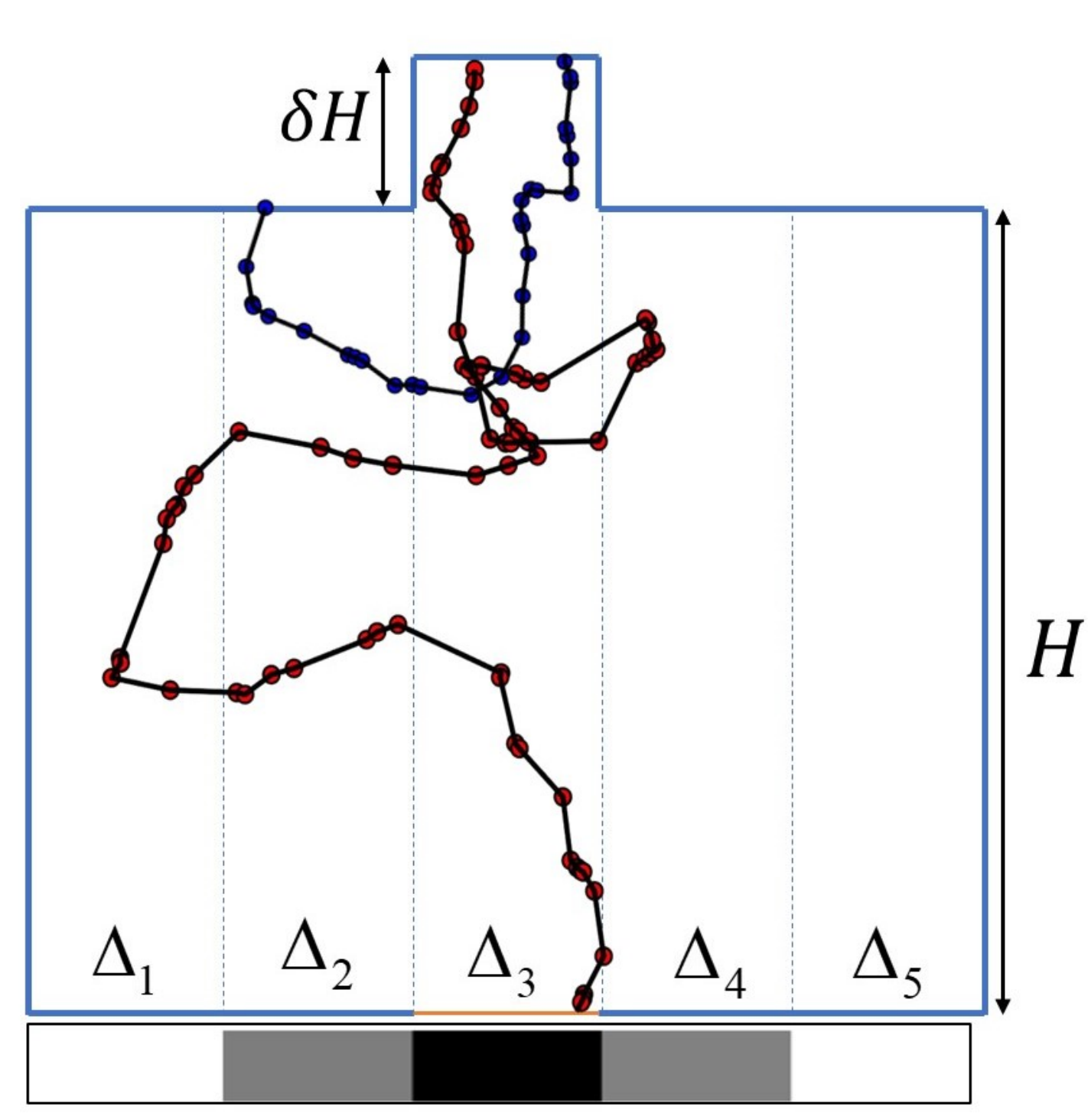}}
	\caption{Schematic representation of the simulated 2D clouds with a bump. The thickness of the cloud $H$, and the bump is $\delta H$. Five different intervals have been considered for the light exit, $\Delta_3$ being right under the bump position.}
	\label{fig:lightscatteringHeight}
\end{figure}

	\begin{figure*}
	\begin{subfigure}{0.49\textwidth}\includegraphics[width=\textwidth]{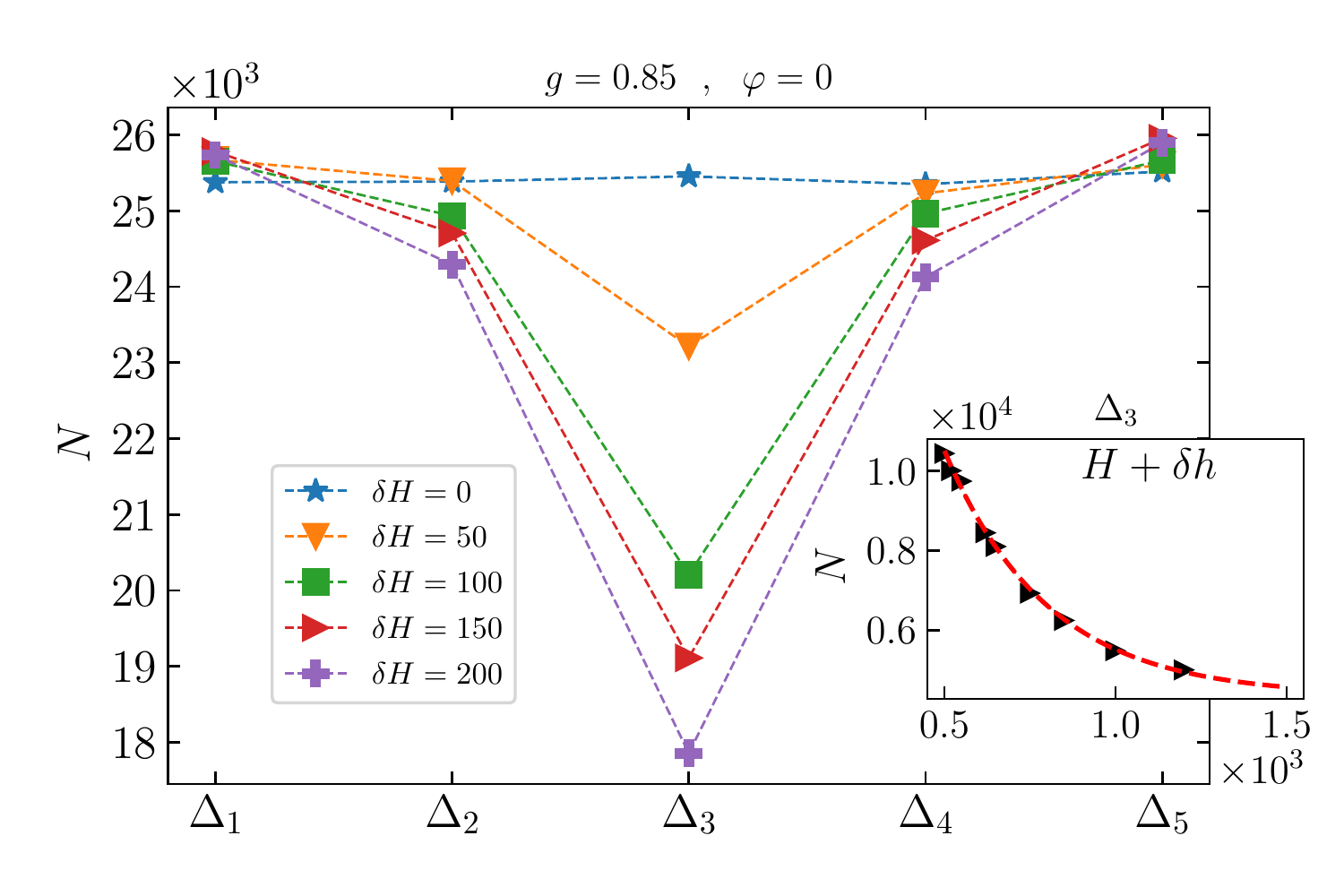}
		\caption{}
		\label{fig:N-delta1}
	\end{subfigure}
	\begin{subfigure}{0.49\textwidth}\includegraphics[width=\textwidth]{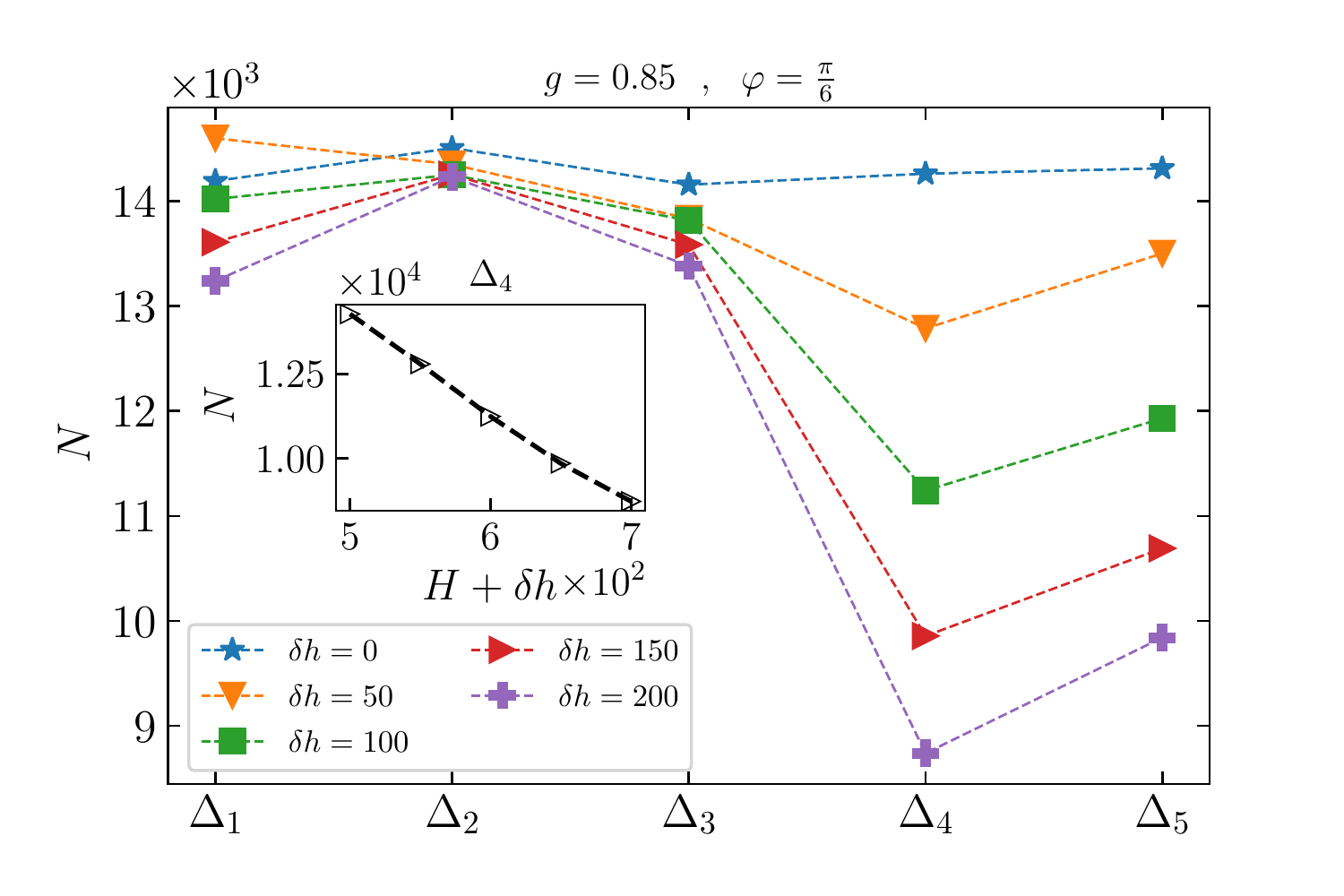}
		\caption{}
		\label{fig:N-delta2}
	\end{subfigure}
	\caption{(a) Output light intensity according to the intervals specified in Fig.~\ref{fig:lightscatteringHeight}, for different $\delta H $ values with $g = 0.85$ for (a) vertical incidence, i.e. angle $\phi = 0$, and (b) non-vertical incidence, i.e. $\phi = \pi / 6$.}
\end{figure*}

The simulations with $g_e = 0.85$ show that the light reaching the bottom of the cloud diminishes exponentially with the cloud thickness, which is rather expected since the probability of scattering at each height is proportional to the intensity of light that reaches that height. A more interesting search is when there is a bump on top of the cloud, which serves as a local thickness fluctuation above the cloud. The light intensity observed in the bottom of the cloud as a function of the bump height gives us valuable information concerning the relationship between the intensity and thickness fluctuations of the cloud. The situation is schematically shown in Fig.~\ref{fig:lightscatteringHeight}, where the height of the bump is $\delta H$. Five intervals have been considered in the bottom of the cloud and only the light beams that exit vertically (with respect to the lower boundary) are counted. The result for the $512\times 512$ system is shown in~Fig.~\ref{fig:N-delta1} and~\ref{fig:N-delta2} for vertical ($\phi=0$) and tilted ($\phi=\frac{\pi}{6}$) incident beams respectively for various $\delta H$ values, and $g=g_e=0.85$. From these figures, one observes that for $\phi=0$ the light intensity reaching the bottom is maximal in the pixel just below the bump, which decays exponentially with $\delta H$, i.e.
\begin{equation}
	N \propto a\exp ( - b(H + \delta H)) + c,
	\label{Eq:ExponentialDecay}
\end{equation}
which supports our definition for the fluctuating field as the logarithm of the intensity, i.e.~\ref{eq:intensity}. $a$, $b$ and $c$ are some constants. For the non-vertical incidence, although the extinction is exponential for all pixels, the pixel with the largest received light intensity is not the one just below the bump position, which is expected since the forward scattering is the probable event in non-zero $g_e$'s. Note also that $g_e$ is generally different for the sunlight and sky blue light, i.e. $g_e (\text{sun}) <g_e (\text{blue})$, indicating that the scattering of sunlight is more isotropic. Given the fact that the more isotropic the scatterings, the lower the sensitivity of the output photons to $\delta H$. Therefore, assuming the same flux, one expects that the fluctuations observed in the photos from the clouds are mainly due to the sky blue light.
Although the validity of this model is very limited, it gives us intuition about the relation between the cloud thickness and the received light. For the cases where the thickness of the cloud is large, the information on the roughness of the top of the cloud is lost to the light rays that come out of the cloud from the bottom. In such situations, the intensity of light beams which reach the bottom of the cloud can be considered to be uniform, and the fluctuation of the received light is due to the fluctuations in the lower boundary (bottom) of the cloud. This can be understood easily by the same arguments like the above, this time for the columns in the lower boundary of the cloud (and the fact that the light reaching the bottom is uniform). Its also notable that, for a better analysis, one has to consider the map that is used in the camera (the lens projection) that builds a 2D photo, see~\cite{Cheraghalizade2020cloud} for details.

\section{Kondev Hyper-Scaling Relations for Self-affine cumulus clouds}\label{SEC:Kondev}
	\begin{figure*}
	\begin{subfigure}{0.5\textwidth}\includegraphics[width=\textwidth]{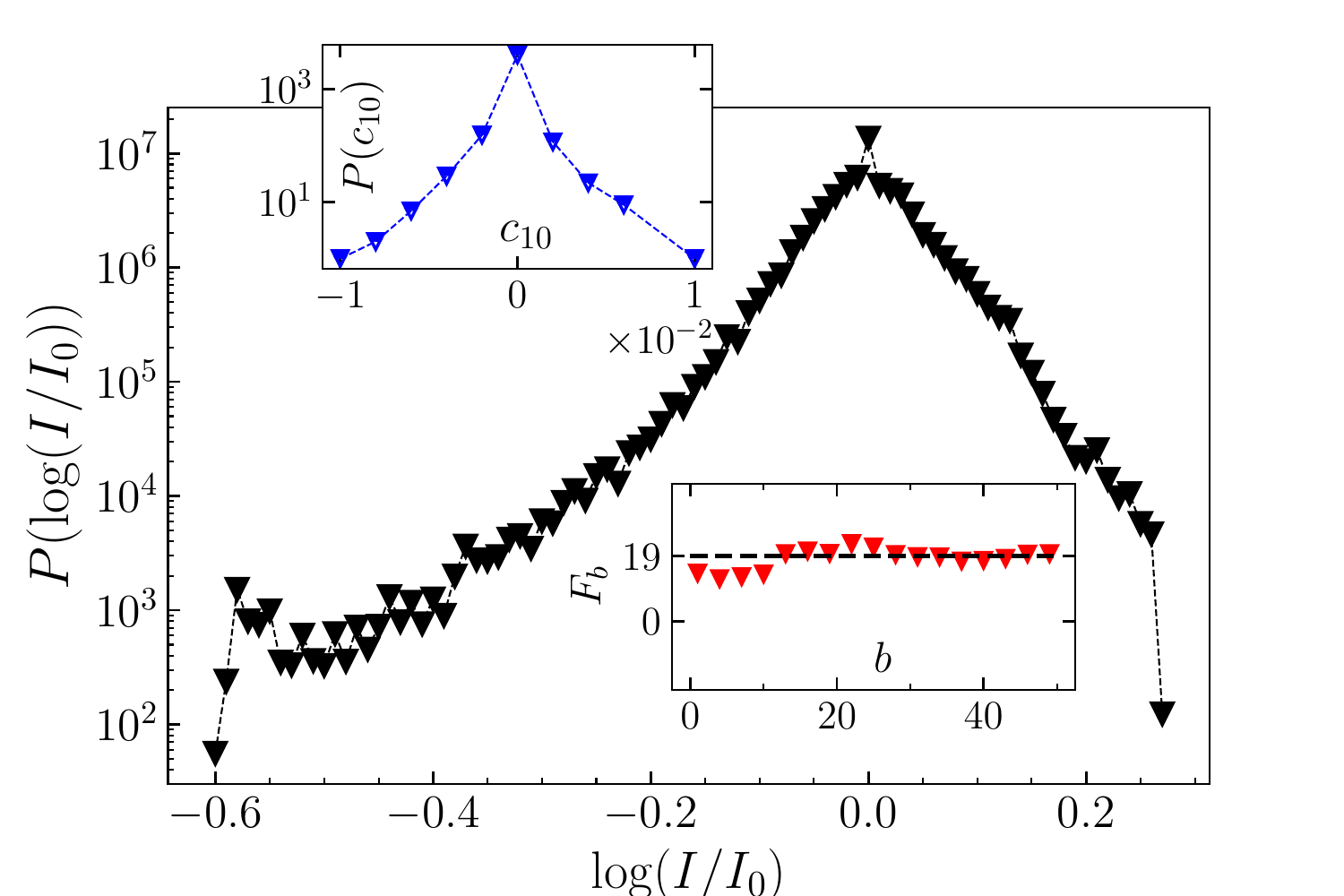}
		\caption{}
		\label{fig:p_h}
	\end{subfigure}
	\begin{subfigure}{0.49\textwidth}\includegraphics[width=\textwidth]{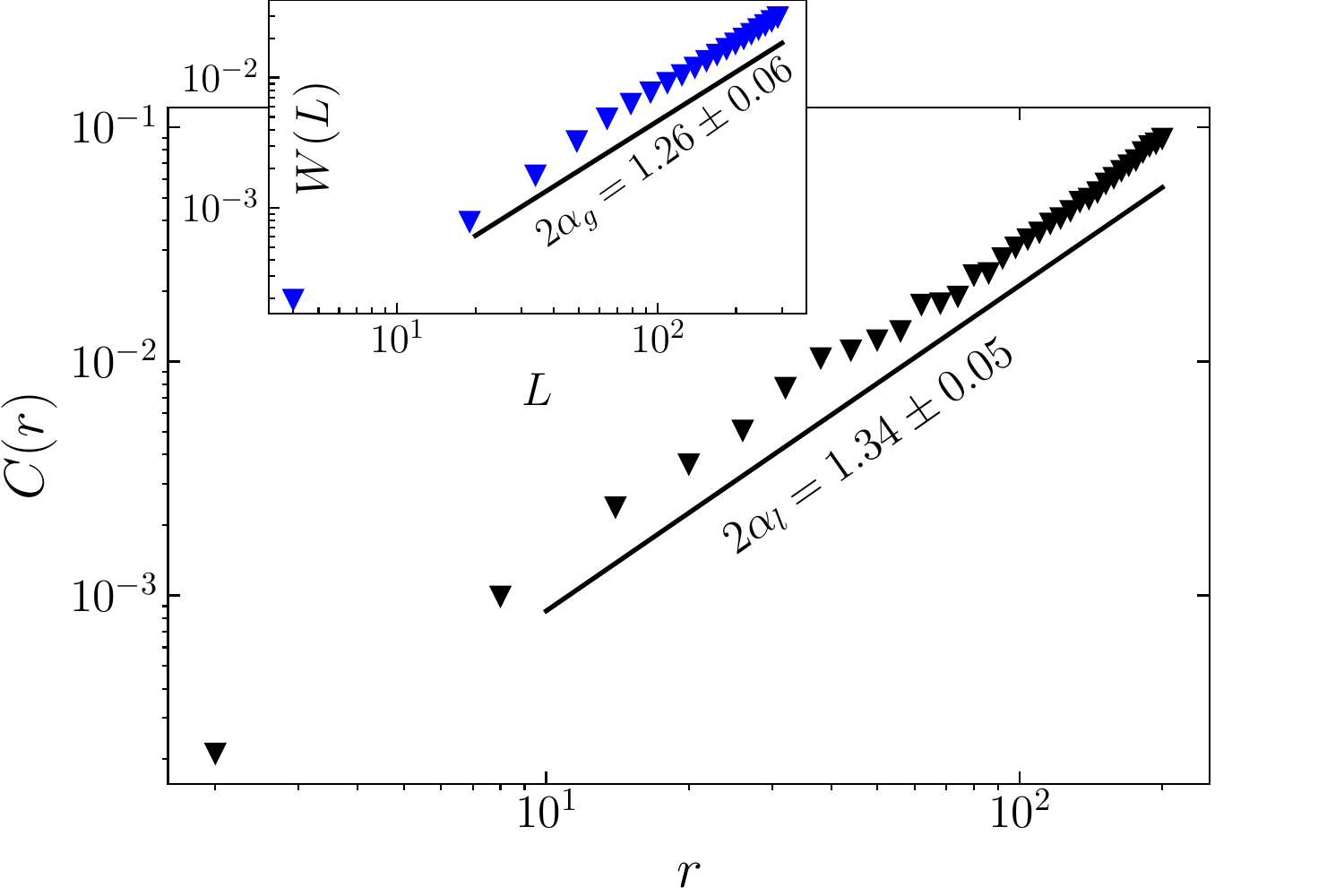}
		\caption{}
		\label{fig:Xinew}
	\end{subfigure}
	\caption{(a) The distribution function of logarithm of the light intensity. Top inset: the distribution function for the local curvature, Lower inset: The kurtosis Eq.~\ref{Eq:Biner} in terms of $b$. (b) The behavior of the local roughness $C (r)$ in terms of $r$ and the global roughness $W$ in terms of the size of the box $L$.}
\end{figure*}

\begin{figure}
	\centerline{\includegraphics[scale=.55]{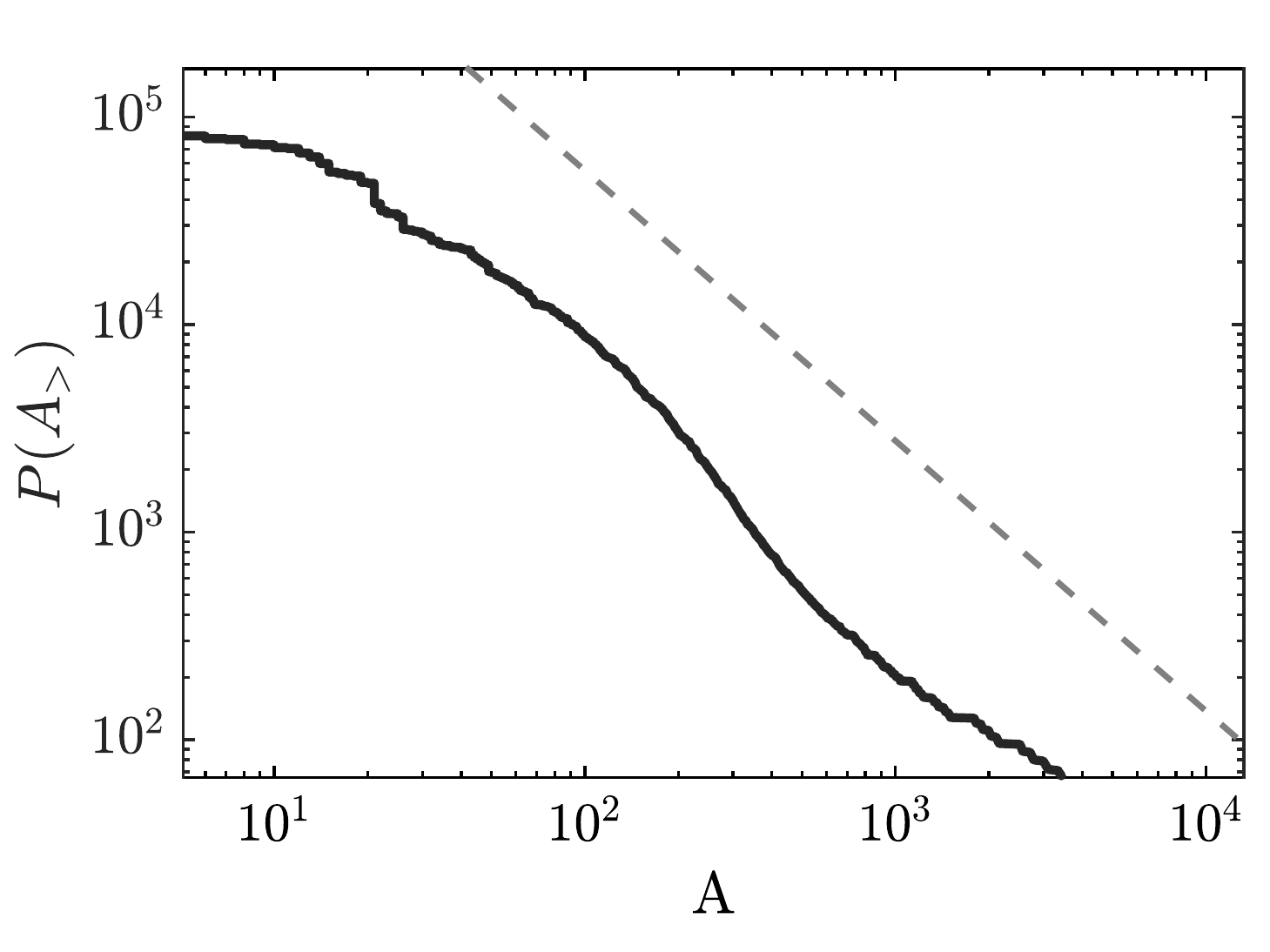}}
	\caption{The distribution function of the area inside the loops $P (A_{>})$. The dashed line is $A^{-1.2}$, corresponding to $\xi=2.4$.}
	\label{fig:Ng}
\end{figure}
In this section we investigate the fractal properties of the cumulus clouds by analyzing the photos. First observe that the distribution function of the normalized intensity deviates from Gaussianity in Fig.~\ref{fig:p_h}. This function is almost linear for small $f$ values, and deviates from the linearity for larger values. Non-Gaussian self-affine rough surfaces have also been observed for different systems, like Graphene~\cite{najafi2017scale,najafi2016conformal}, diffraction theories~\cite{zhao1997diffraction} and Pearson system of frequency curves~\cite{zhan2019modeling,belhadjamor2020numerical}, non-Gaussian height distribution was modeled using the Weibull probability distribution~\cite{sabino2022impact}, rough Surface Morphologies~\cite{kondev2000nonlinear}. Our analysis adds cumulus clouds to this list, stating that when they are mapped to non-Gaussian rough surfaces. The other measures, like the distribution function of local curvature (see inset of Fig.~\ref{fig:p_h}), and also the Kurtosis (the lower inset of Fig.~\ref{fig:p_h}) confirm this conclusion. No matter a self affine rough surface is Gaussian or non-Gaussian, the roughness exponent is well-defined for them using the relations Eq.~\ref{Eq:roughness} and Eq.~\ref{Eq:correlation}. These functions are shown in Fig.~\ref{fig:Xinew}, from which we see that $ \alpha_l=0.67 \pm 0.05 $ and $ \alpha_g=0.63 \pm 0.05 $. These exponents are the same given their error bars, revealing that the surgace is mono-fractal.\\

In order to test the Kondev relations Eqs.~\ref{Eq:KondevTau_r},~\ref{Eq:KondevTau_l},~\ref{Eq:KondevX_l},~\ref{Eq:KondevHyper}, and~\ref{Eq:Hyper2}, we use the data that obtained by analyzing the level lines in the Ref.~\cite{najafi2021self}. The fractal dimension of boundary contour line was find to be $D_f=1.248 \pm 0.006$, and the fractal dimension of bulk contour line, which is $D_f=1.22 \pm 0.02$, and the exponents for the distribution function for radius of gyration $r$ and the loop length $l$, which reported to be $\tau_r=2.12 \pm 0.03$  and $\tau_l=2.38 \pm 0.02$, respectively. Also the green function were shown to be logarithmic, giving rise to $x_l=0$. In the present paper we calculated the function $N_> (A) $ the exponent of which, defined by $ N_> (A) \sim A^{-\xi/2}$, is estimated to be $\xi \cong 2.4 $ (see Fig.~\ref{fig:Ng}). Given this information, one calculated the Kondev's hyper-scaling relation, which is reported in table~\ref{tab:exponents}. From this figure we see that all of the Kondev hyper-scaling relations are violated, other than Eq.~\ref{Eq:KondevHyper}, which has its roots in the fact that the conditional probability $p(l|r)$ is a narrow function of both $l$ and $r$. This proves that the rough surface obtained by mapping the cumulus clouds is an unconventional non-Gaussian self-affine rough surfaces. 
\begin{table*}
	\begin{tabular}{c | c | c | c | c | c | c }
	\hline  & Eq.~\ref{Eq:KondevTau_r} & Eq.~\ref{Eq:KondevTau_r} & Eq.~\ref{Eq:KondevTau_l} & Eq.~\ref{Eq:KondevX_l} & Eq.~\ref{Eq:KondevHyper} & Eq.~\ref{Eq:Hyper2}\\
		\hline The Equation & $\frac{\tau_r}{1+D_f(\tau_l-1)}$ & $\frac{\tau_r}{3-\alpha}$ & $D_f\frac{\tau_l-1}{2-\alpha}$ & $D_f\frac{\tau_l-3}{2x_l-2}$ & $D_f\frac{\tau_r-1}{\tau_l-1}$ & $\frac{x_l}{2-\alpha}$\\
	\hline Value & $0.79\pm 0.02$ & $0.90\pm 0.02$ & $1.24\pm 0.05$ & $0.38\pm 0.01$ & $0.99\pm 0.03$ & $1.77\pm 0.16$ \\
	\hline validity & $\times$ & $\times$ & $\times$ & $\times$ & \checkmark & $\times$ \\
		\hline
	\end{tabular}
	\caption{This table shows the validity of the hyper-scaling relations. The signs $\times$ and \checkmark in the last line shows that the hyper-scaling relation is invalid/valid respectively.}
	\label{tab:exponents}
\end{table*}

\section{Conclusion}
In this paper we mapped the fractal cumulus clouds to rough surfaces. To this end we analyzed the photos, taken from cumulus clouds in Ardabil, Iran in June-July 2018. In the first part of the article, after reviewing the standard theory of rough surfaces, and also exploring a coarse-grained phenomenological model of light scattering of cumulus clouds, we assessed some previously established claims for the fractal properties of cumulus clouds. This model gives effective $g$ parameter, defined in Eq.~\ref{Eq:P}. We showed that there is a connection between the cloud thickness, and the logarithm of the received visible light from them. In particular, by analyzing the results of simulations, we argued that the intensity of the light reaching the ground exponentially depends on the thickness of the cloud protrusion just above the target area. This explains why we are using logarithm of the intensity as the main fluctuating field.\\

In the second part of the paper we concentrated on the statistical properties of the logarithm of the light intensity received from the cumulus clouds. From this analysis, we asserted that the resulting rough surface is self-affine, with some scaling relations for the observables like the systems roughness, and the distribution of the fluctuating filed. Our results suggest that the system is mono-fractal in the sense that the global and local roughness exponents are the same, i.e. $\alpha= \alpha_g = \alpha_l$. By analyzing various distribution function, we numerically showed that the system is non-Gaussian self-affine surface, with anomalous distribution functions. The other anomalous behavior of the system is concerning the hyper-scaling relations between critical exponents of conventional self-affine surfaces (Kondev relations) which are violated in our case (Table~\ref{tab:exponents}).

\bibliography{refs}
	
\end{document}